# FRICTION JOINTING OF DISTRIBUTED RIGID CAPACITORS TO STRETCHABLE LIQUID METAL COIL FOR FULL-BODY WIRELESS CHARGING CLOTHING

*Takashi Sato[1], Shinto Watanabe[2], Ryo Takahashi[3], Wakako Yukita[3], Tomoyuki Yokota[3], Takao Someya[3], Yoshihito Kawahara[3], Eiji Iwase[2], and Junya Kurumida[1]*

[1] National Institute of Advanced Industrial Science and Technology, JAPAN,

[2]Waseda University, JAPAN, and [3]The University of Tokyo, JAPAN

## ABSTRACT

For full-body wireless power transfer (WPT), a liquid metal (LM)-based meandered textile coil has been proposed. Multiple rigid capacitors must be inserted in a long coil for efficiency; however, the conventional adhesive jointing suffers from the fragile connection between a rubber tube filled with LM and the capacitor due to the poor adhesion of the rubbers. This paper presents a friction-based jointing, which covers the capacitor with a rigid capsule to enhance the frictional force between the tube and capsule. By experimentally optimizing the capsule design, the LM coil with capacitors showed 3.1 times higher stretch tolerance (31.8 N) and 3.5 times higher bending tolerance (25.9 N) than the adhesive jointing. Moreover, the WPT garment prototype shows excellent mechanical durability against repeated stretching and washing over 100 times. Our full-body meandered textile coil can enable wireless charging to wearable devices around the body for long-term continuous healthcare monitoring, activity recognition, and AR/VR.

## INTRODUCTION

Wearable sensing networks consisting of motion trackers, healthcare sensors, etc., can continuously monitor full-body human physiological signals[1],[2]. Body-scale WPT clothing with efficiency and stretchability is promising for long-term continuous operation of multiple wearable devices placed freely around the body. A meandered textile coil[3],[4], which consists of a zigzag coil pattern and a low-loss LM-based stretchable tube, enables safe, body-scale, watt-class WPT by generating a strong inductive field only near the skin surface. To design such a body-scale coil at high frequency, multiple capacitors must be inserted into the long coil to avoid stray impedance issues[5]. However, the connections between the rigid capacitor and the rubber tube filled with LM break when stretched, owing to the poor adhesion of rubbers such as silicone and urethane. This paper presents a friction-based jointing method; a rigid capacitor covered with a rigid capsule is inserted into a silicone rubber tube, and the tube is fixed by the frictional force between the capsule and tube. The capsule enhances the frictional force and suppresses the breakage of the tube around the edge of the rigid capacitor.

## METHODS

Figure 1 shows the schematics of the WPT clothing. As shown in Figures 1(a) and (b), the LM-based transmitter (TX) coil is placed around the clothing and supplies power to the receiver (RX) coil around the body using magnetic resonant coupling (MRC). Figures 1(c) and (d) show the distributed capacitors to a meander-shaped LM coil. Figure 2 shows the schematics of the LM meandered coil with distributed capacitors. Figure 2(a) shows configurations of the WPT garments; the LM coil is inserted into the pocket of the machine-knitted WPT clothing. This study proposes a jointing method using frictional force, compared to the conventional jointing method using adhesive force, as shown in Figures 2(b) and (c). Since the frictional force is attributed to both the pressure and contact area, the effects of the diameter and length of the capsule on the stretch and bending tolerances of the coil are investigated.

Figure 3 shows the cross sections and photographs of a sample for the stretch and bending tests of the frictional jointing method. As shown in Figures 3(a) and (b), a UV-cured 3D-printed cylindrical jig (Max X 27, Asiga, Australia) imitating the capsule is inserted into a silicone rubber tube. The tip of the jig is conical shape for insertion and is separated after insertion to release the stress concentration around the tips, as shown in Figure 3(c). To prevent the tube from rupturing, the inner and outer diameters of the tubes are 2 and 3 mm, respectively. The outer diameter of the jig ($\varphi_{jig}$) is from 1.6 to 8.0 mm, the inner diameter of the jig is 1.0 mm, and the length of the jig ($L_{jig}$) is from 2 to 30 mm. The cone apex angle is 30°, and the cone tip has a radius of curvature of 0.5 mm. The separation area is 0.2 mm thick. The tube is pulled parallel and perpendicular to the jig in the tensile and bending tests, respectively, to investigate the frictional force and breakage of the jig and tube.

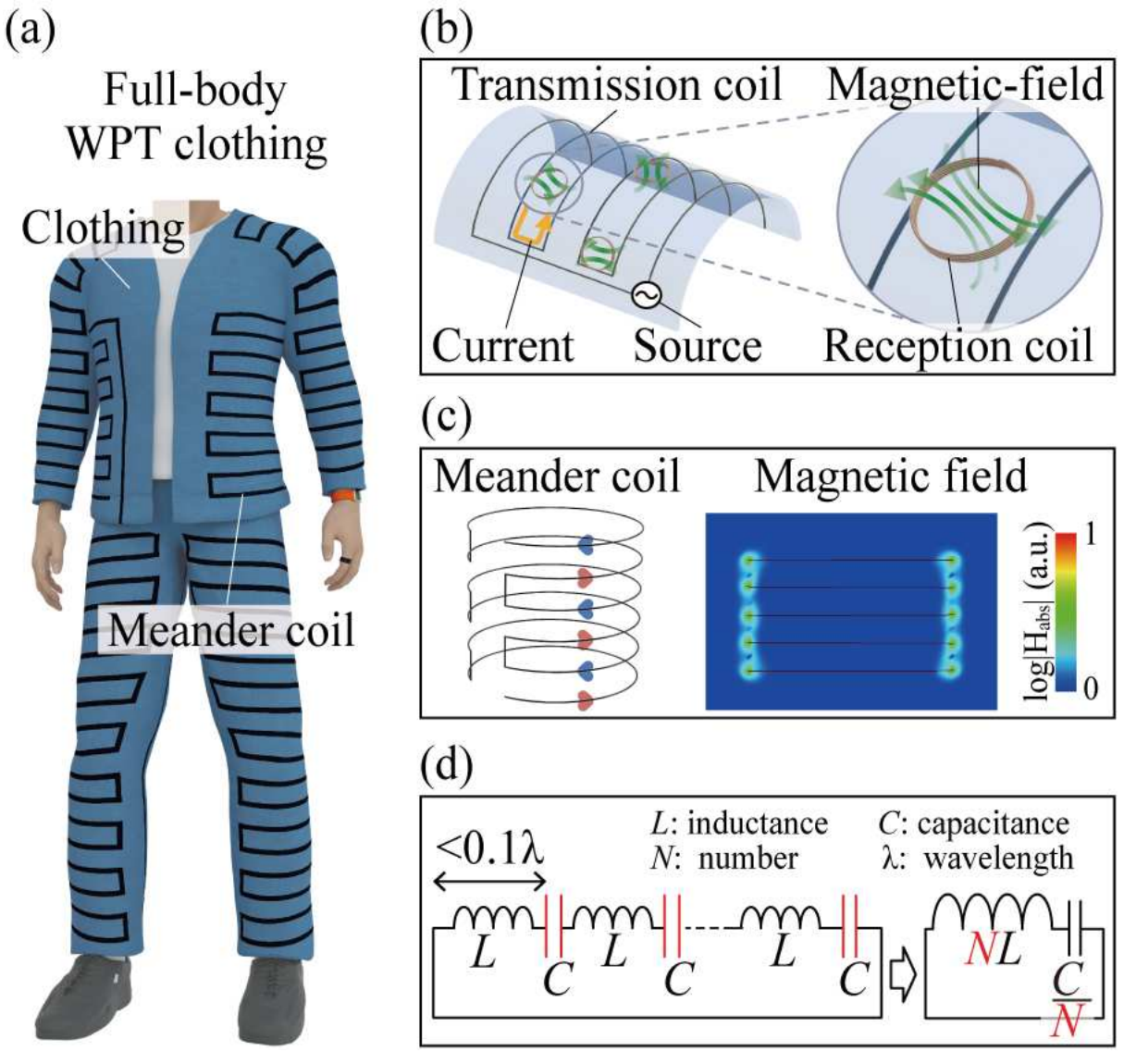

*Figure 1 Schematics of full-body WPT clothing: (a) Overview, (b) WPT systems, (c) Meander coil design, (d) Distributed arrangement of the capacitance.*

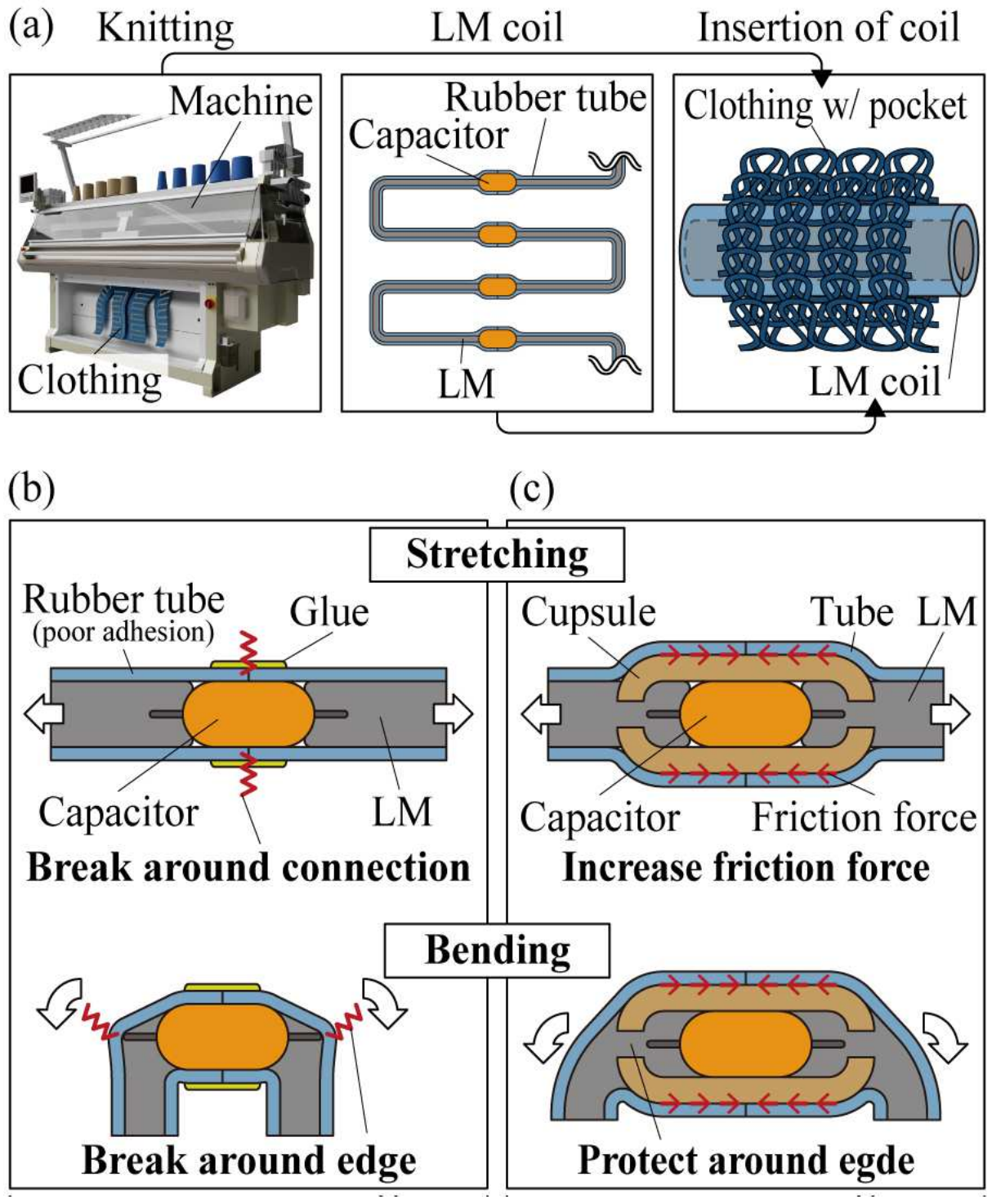

*Figure 2 Schematics of the LM-based meander coil with distributed capacitors: (a) Configuration of the WPT clothing, (b) Conventional jointing method using adhesive force, (c) Proposed jointing method using friction force.*

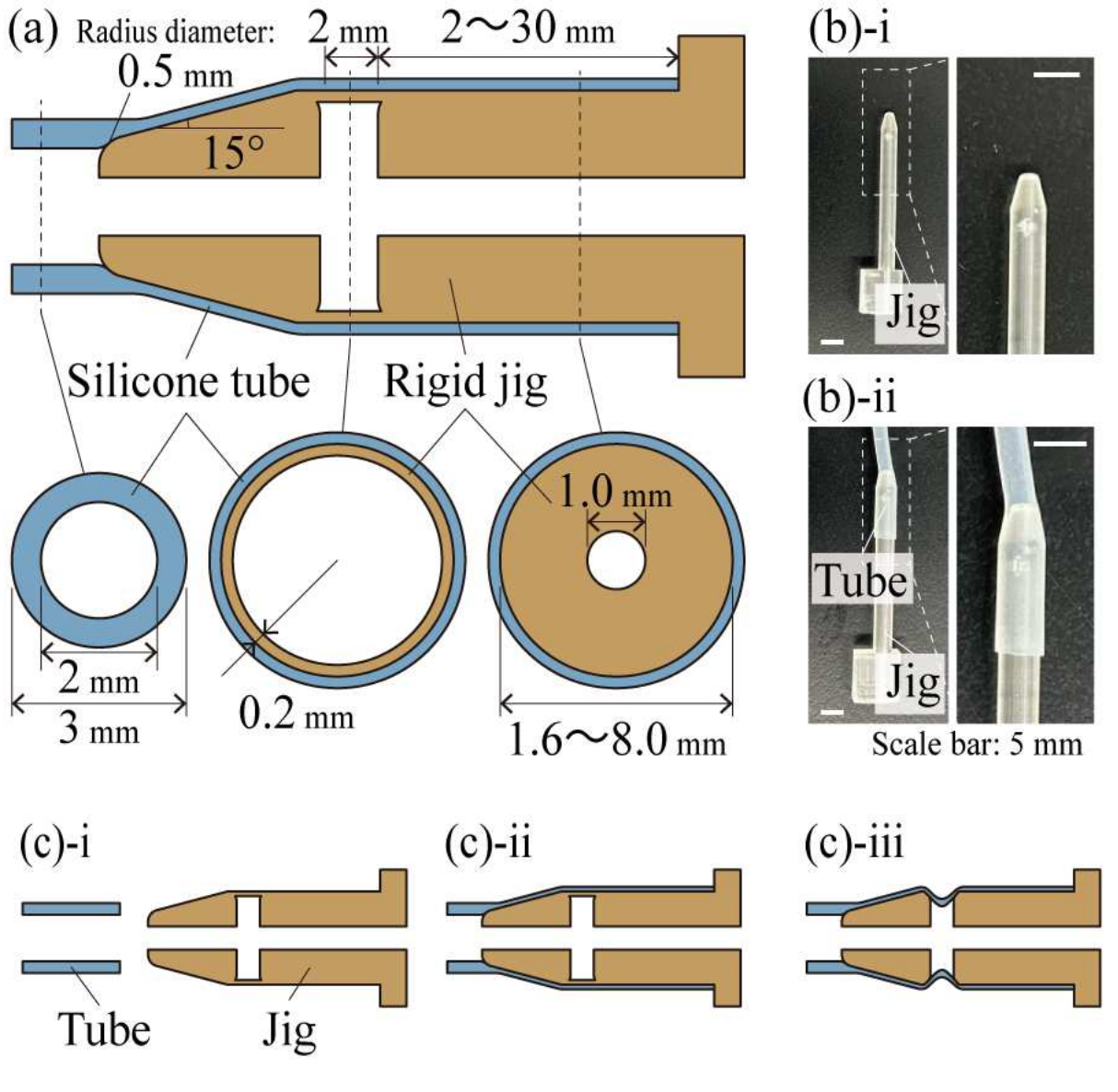

*Figure 3 Experimental samples for stretching and bending tests of friction-based jointing: (a) Cross sections, (b) Photographs, (c) Fabrication procedures.*

Figure 4 shows the cross sections and photographs of the frictional jointing of a capacitor and LM-based coil. A leaded capacitor (AxiMax 400 Comm C0G, KEMET, USA) covered with a rigid capsule is jointed with rubber tubes filled with galinstan (Sichuan HPM, China), as shown in Figures 4(a) and (b). Galinstan has high conductivity ($10^6$ S/m) and stretchability (>400%)[6],[7]. A capacitor with a thin rubber tube was covered with a capsule and inserted into the rubber tubes. The ends of the tubes are sealed with adhesive glue to prevent leakage. The outer diameter ($R_{capsule}$) and length ($L_{capsule}$) of the capsule are 4.8 and 36 mm, respectively, based on the measurement results. Figure 4(c) shows fabrication procedures; tubes filled with galinstan and the capacitor covered with the capsule are assembled, and the tips of the capsules are separated.

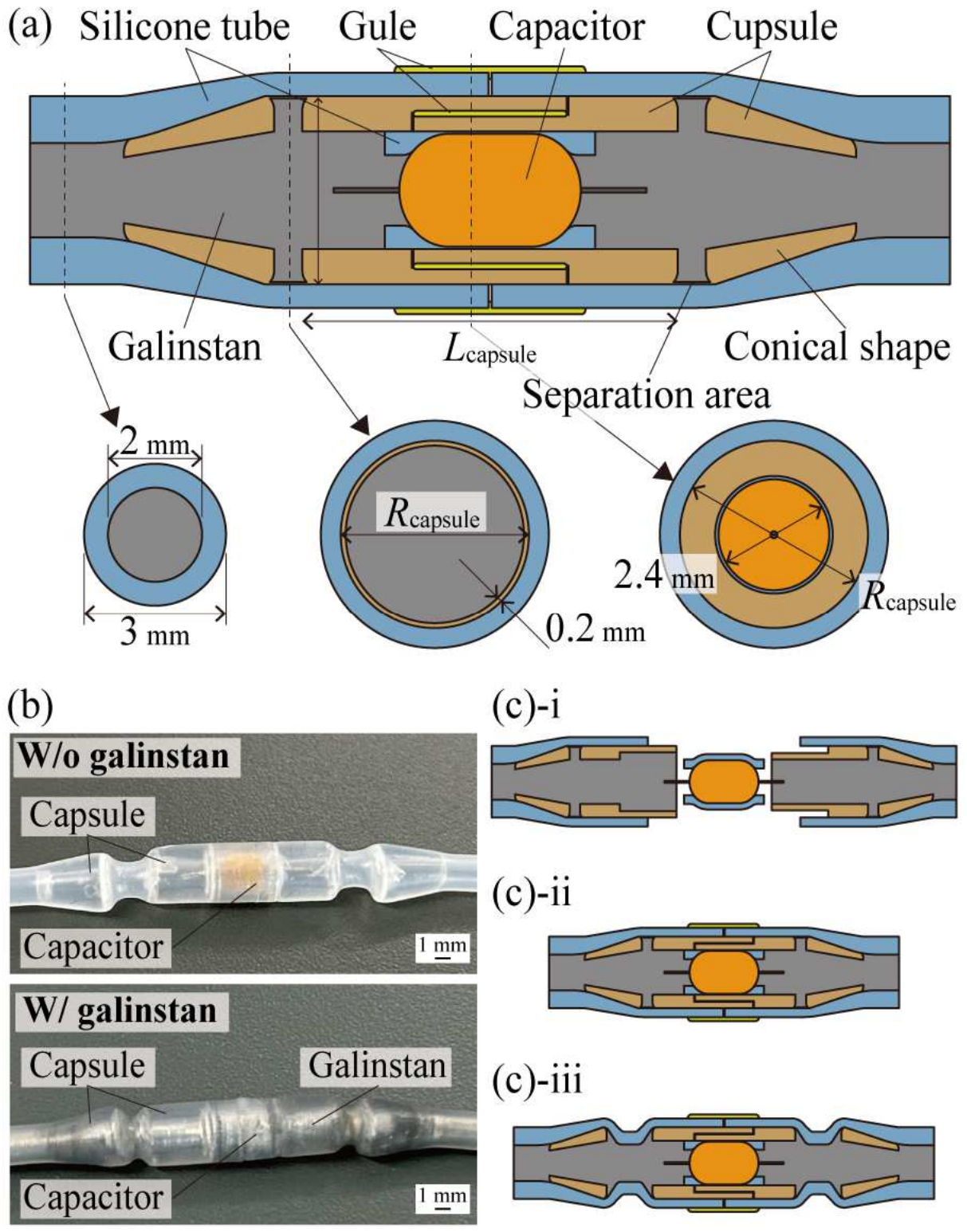

*Figure 4 Friction-based jointing between capacitors and rubber tube filled with galinstan: (a) Cross sections, (b) Photographs, (c) Fabrication procedures.*

# RESULTS AND DISCUSSION

Figure 5 shows the measurement results of the stretch tolerance. As $\varphi_{jig}$ and $L_{jig}$ increased in Figures 5(a) and (b), the tube slipped off from the jig and the breakage force ($F_{break}$) increased. As $\varphi_{jig}$ and $L_{jig}$ increased further, the tube ruptured and $F_{break}$ became constant. These results indicate that the frictional force increased and the stretch tolerance improved as $\varphi_{jig}$ and $L_{jig}$ increased. When sweeping $\varphi_{jig}$ and $L_{jig}$, $F_{break}$ increased and became constant as $\varphi_{jig}$ and $L_{jig}$ increased. The area surrounded by the red dotted line in Figure 5(c) shows the values of $\varphi_{jig}$ and $L_{jig}$, where $F_{break}$ was constant. These results indicate that the highest stretch tolerance can be obtained by increasing $\varphi_{jig}$ and $L_{jig}$ until the stretch tolerance of the tube is brought out.

Figure 6 shows the measurement results of the bending tolerance. For small $\varphi_{jig}$ and $L_{jig}$, the tube slipped off from the jig. As $\varphi_{jig}$ increased in Figure 6(a), the jig broke and $F_{break}$ increased. As $\varphi_{jig}$ increased further, the tube ruptured and $F_{break}$ decreased. As $L_{jig}$ increased in Figure 6(a), the tube ruptured and $F_{break}$ increased. As $L_{jig}$ increased further, the jig broke and $F_{break}$ decreased. These results indicate that the frictional force increased and the bending tolerance improved as $\varphi_{jig}$ and $L_{jig}$ increased. Within the range where the tube did not slip off, we found a trade-off relationship: the suppression of the breakage of the jig and the promotion of the difference in the bending stiffness between the jig and tube by using a thick and short jig. When sweeping $\varphi_{jig}$ and $L_{jig}$ in Figure 6(c), $F_{break}$ increased and then decreased as $\varphi_{jig}$ and $L_{jig}$ increased, respectively. This trade-off relationship can be overcome and high bending tolerance can be achieved by increasing the toughness and flexibility of the capsule.

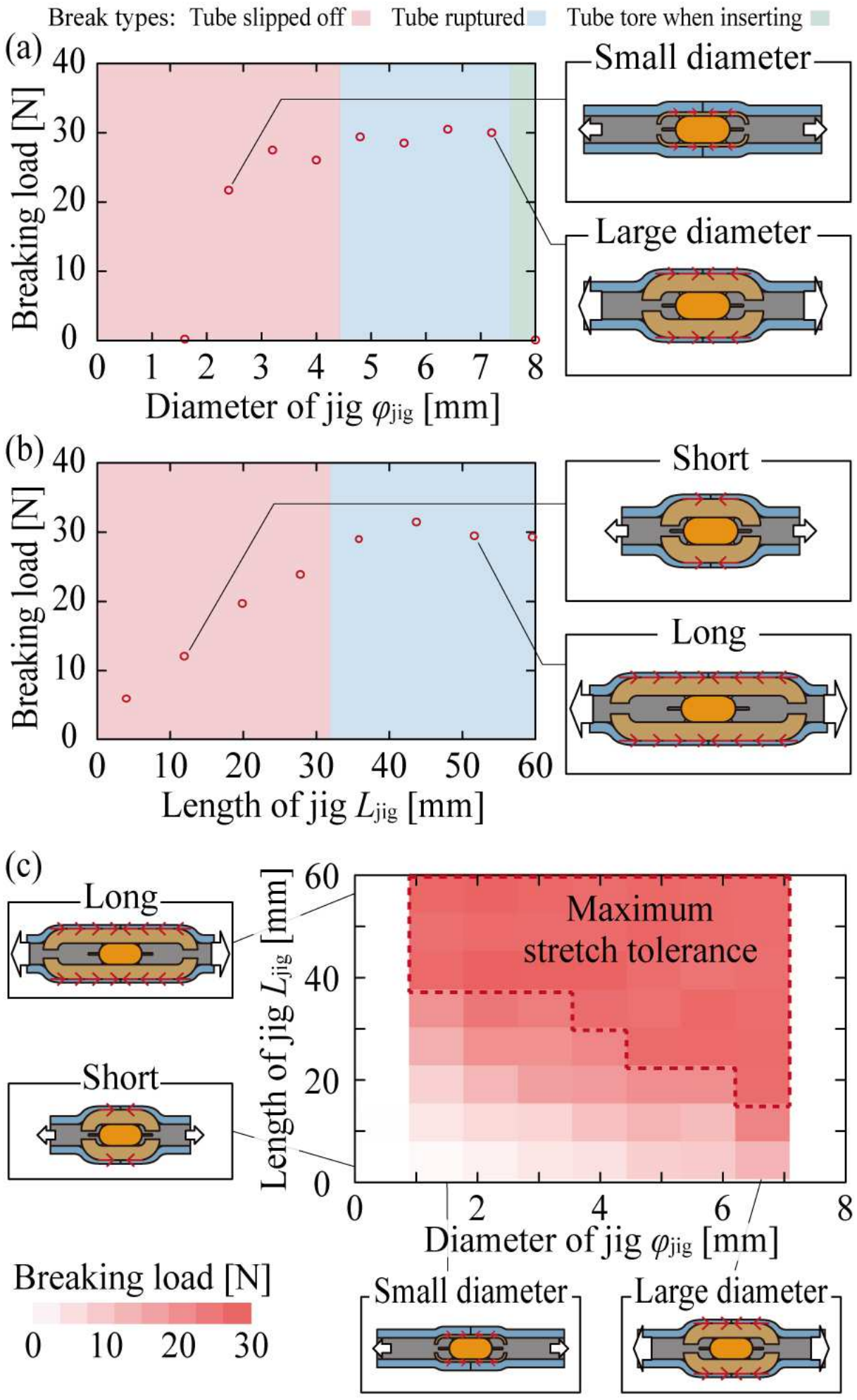

*Figure 5 Experimental results of stretch tolerance: (a) Relationship between $\varphi_{jig}$ and $F_{break}$ breaking load ($L_{jig}$=18 mm, N=3), (b) Relationship between $L_{jig}$ and $F_{break}$ ($\varphi_{jig}$=4.8 mm, N=3), (c) Mapping of $F_{break}$ for $\varphi_{jig}$ and $L_{jig}$ (N=3).*

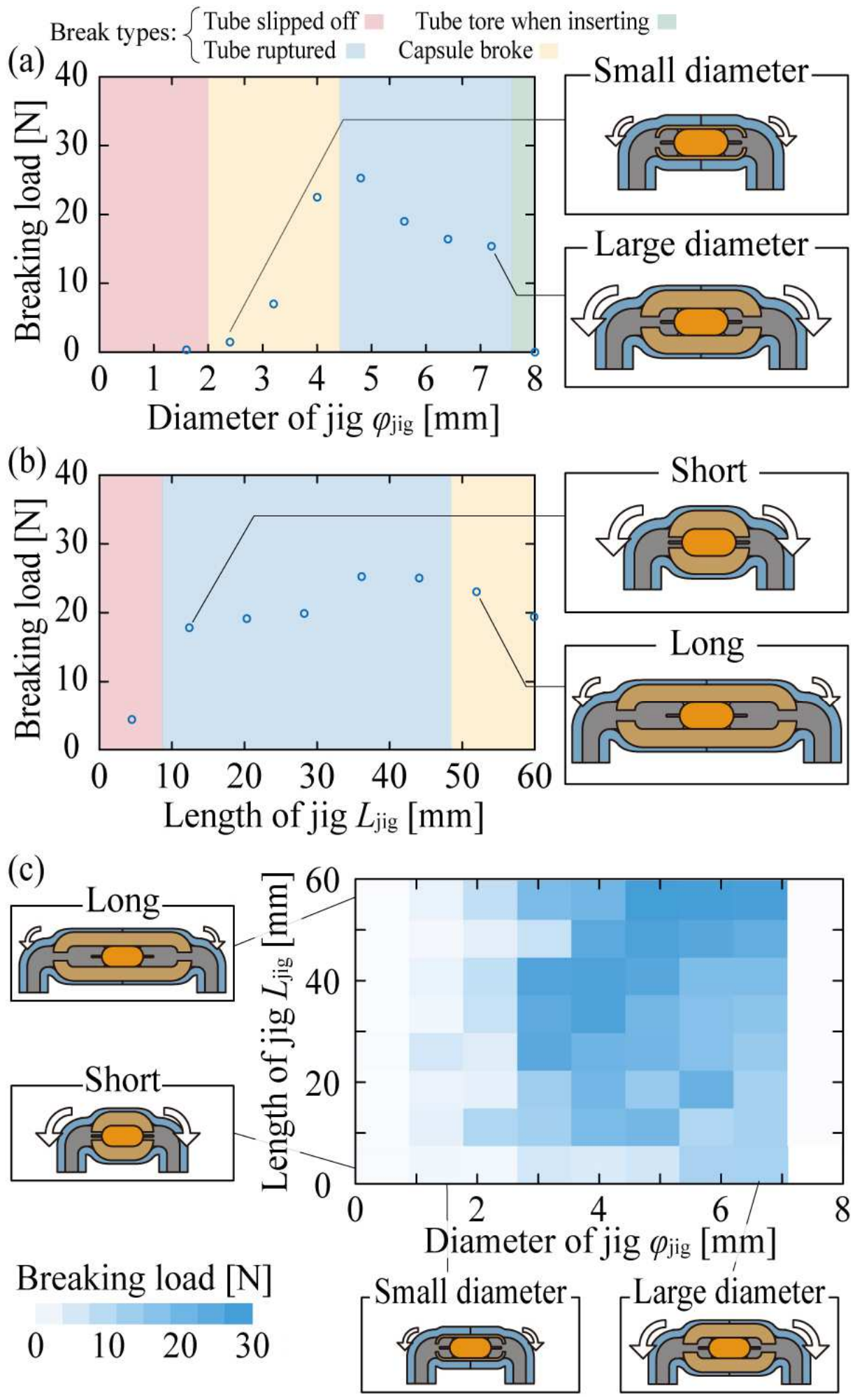

*Figure 6 Experimental results of bending tolerance: (a) Relationship between $\varphi_{jig}$ and $F_{break}$ ($L_{jig}$=18 mm, N=3); (b) Relationship between $L_{jig}$ and $F_{break}$ ($\varphi_{jig}$=4. mm, N=3), (c) Mapping of $F_{break}$ for $\varphi_{jig}$ and $L_{jig}$ (N=3).*

Figure 7 compares the stretch and bending tolerances of the jointing methods. Figure 7(a) shows the product of the stretch and bending tolerances normalized by their maximum values. The maximum stretch and bending tolerances were31.8 and 25.9 N when $\varphi_{jig}$ and $L_{jig}$ were 4.8 and 18 mm, respectively. For comparison, stretch and bending tolerances were measured for conventional adhesive glues: Aron Alpha Extra Jelly (TOAGOSEI, Japan), 1530C (ThreeBond, Japan), Super X (Cemedine, Japan), and Ultra-Versatile SU (KONISHI, Japan). Their average stretch and bending tolerances were 10.3 and 7.5 N, respectively as shown in Figures 7(b) and (c). The frictional jointing method achieved 3.1 times higher stretch tolerance and 3.5 times higher bending tolerance than the adhesive jointing method.

Figure 8 shows photographs of the WPT clothing by the friction jointing method. Figure 8(a) shows the Prototype of the WPT cardigans, sleeves, and trousers. The LM-based TX coil for the cardigan has eight capacitors of 270 pF attached at 90 mm intervals. The TX coil for the sleeve has six capacitors of 470 pF attached at 50 mm intervals. The TX coil for the trousers has 14 capacitors of 390 pF attached at 90 mm intervals. Their resonant frequency is adjusted to 6.78 MHz, the same as the AC source. The conductivity of the LM coil is $10^{-1}$ Ω/m and its weight is 27.8 g/m. The length and width of the RX coil are 20 mm and its resonant frequency is adjusted to 6.78 MHz. The inductance, resistor, and Q-factor of the RX coil are 3.3 μH, 2.3 Ω, and 61, respectively. Figure 8(b) demonstrates WPT for five devices with multiple LEDs placed on the cardigan, which requires Watt-class power supply. This result shows that watt-class WPT can be achieved for high-power wearable devices using our WPT clothing. Figure 8(c) shows photographs of the bending test. The sleeve was attached to the elbow and was repeatedly bent. The LM coil inside the sleeve maintained its electrical connection after 100 bends. Figure 8(d) shows photographs of the washing test. The cardigan was washed repeatedly using a washing machine. The impedance of the coil inside the cardigan increased by 1.7% after 100 washes. These results indicate that WPT clothes are highly durable against everyday mechanical deformation.

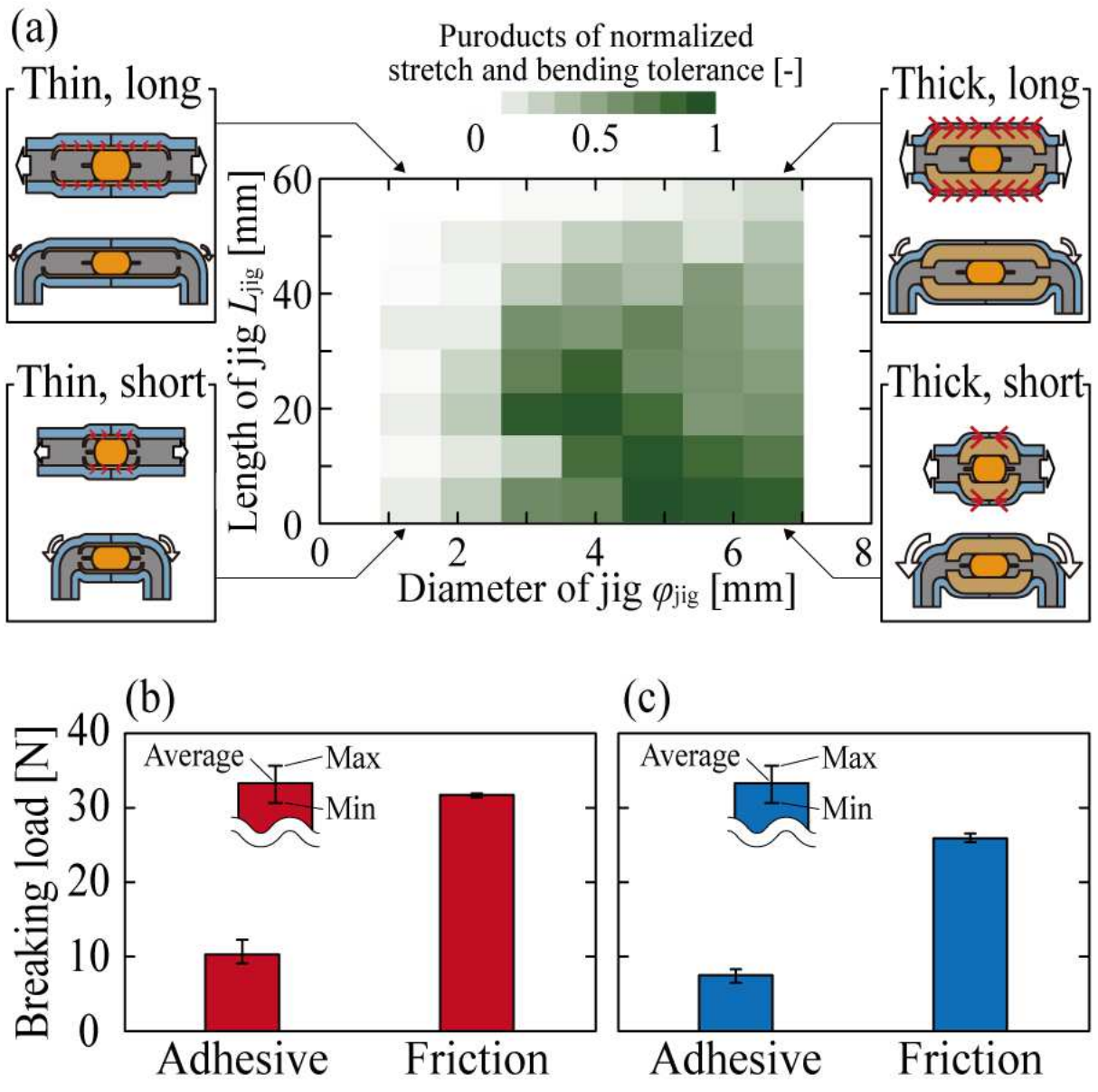

*Figure 7 Comparison of stretch and bending tolerances for jointing methods: (a) Stretch and bending tolerances for the friction jointing method, (b) Stretch and (c) Bending tolerances for jointing methods (N=3).*

## CONCLUSIONS

This study proposes an LM-based meandered coil with distributed capacitors using the friction jointing method for the full-body WPT clothing and demonstrates its stretch and bending tolerance. While the stretch tolerance was mainly affected by the frictional force, the bending tolerance was affected by the frictional force, the bending stiffness of the capsule, and the difference in the bending stiffness between the capsule and tube, and these factors were in a trade-off relationship. By optimizing the capsule shapes experimentally, the frictional jointing method achieved 3.1 times higher stretch tolerance and 3.5 times higher bending tolerance than the conventional adhesive jointing method. Furthermore, we fabricated the prototype of WPT clothing, cardigans, sleeves, and trousers, and demonstrated WPT and repeated stretching and washing. Our full-body WPT technology can be integrated into various textile products and contribute to the development of sustainable wearable computing in everyday life.

## ACKNOWLEDGEMENTS

This work was partly supported by the JST ACT-X JPMJAX21K6, the JST ACT-X JPMJAX21K9, and JSPS KAKEN 22K21343.

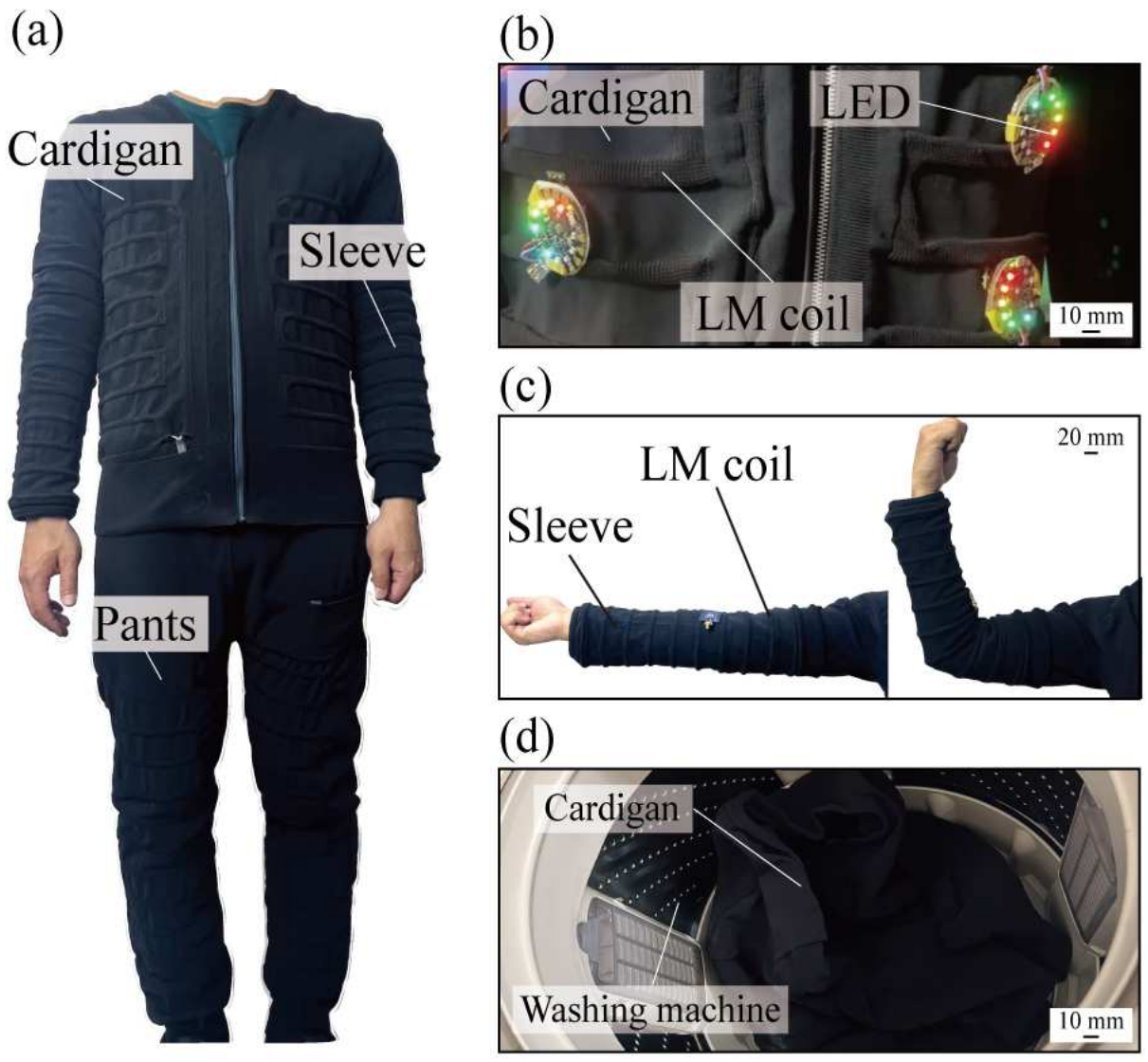

*Figure 8 Photographs: (a) Prototype of the WPT clothing, (b) LED testing using WPT, (c) Bending tests, (d) Washing tests.*

## CONTACT

T. Sato tel: +81-80-2185-3561
e-mail: machotakashi-satou@aist.go.jp